\def\plb#1#2#3{{\it Phys.\ Lett.} {\bf B#1}, #2 (#3)}
\def\prd#1#2#3{{\it Phys.\ Rev.} {\bf D#1}, #2 (#3)}

\def\st{\scriptstyle}		
\def\sss{\scriptscriptstyle}	

\newcommand{\real}{\Re\mbox{e}\,}
\newcommand{\imag}{\Im\mbox{m}\,}

\def\thingie{\hbox{\kern-9pt\raise1pt%
         \hbox{{\fiverm(}{\lower1.5pt\hbox{\twelvebf--}}{\fiverm)}}}}
\newcommand{\optbar}[1]{\shortstack{{\tiny (\rule[.4ex]{1em}{.1mm})} 
  \\ [-.7ex] $#1$}}		
\def\pmdiff#1#2{\raise.5ex\hbox{$\sss +#1}$%
    \kern-2.8em\lower1ex\hbox{${\sss-#2}$}} 

\def\barp{{\raise.35ex\hbox{${\sss (}$}} --- {\raise.35ex\hbox{${\sss )}$}}}				
\def\bdbarp{\hbox{$B_d$\kern-1.4em\raise1.4ex\hbox{\barp}}}
\def\nlpbarp{\hbox{$\nu_{\ell^{\prime}}$\kern-1.4em \raise1.4ex\hbox{\barp}}}

\def\ne{\nu_e}
\def\nm{\nu_\mu}
\def\nt{\nu_\tau}
\def\ni{\nu_i}

\def\dm2{\delta M^2_{ m \, m^{\sss\prime}}}

\def\ket#1{|#1\rangle}

\def\decayarrow{\kern0.2em\hbox{$\raise1.08ex\hbox{\big|}\kern-0.5em
                \longrightarrow$}}

\def\gsim{\;\raisebox{-.6ex}{$\stackrel{>}{\sim}$}\;}
\def\pom{\raisebox{-.8ex}{$\stackrel{+}{{\sss (}-{\sss )}}$}}
\def\ra{\rightarrow}

\newcommand{\Eq}[1]{Eq.~(\ref{eq#1})}
\newcommand{\beq}{\begin{equation}}
\newcommand{\eeq}{\end{equation}}

\documentclass{cernrep} 
\usepackage{texnames}
\usepackage[T1]{fontenc}
\begin{document}

\title{Neutrino Oscillation Physics\footnote{FERMILAB-FN-0946-T. Published in the {\em Proceedings of the International School on AstroPaticle Physics}, eds.\ G. Bellini and L. Ludhova (IOS Press, Amsterdam, 2012) p.1.}}
\author{Boris Kayser}
\institute{Fermilab, Batavia, IL USA}
\maketitle             

\begin{abstract}
To complement the neutrino-physics lectures given at the 2011 International School on Astro Particle Physics devoted to Neutrino Physics and Astrophysics (ISAPP 2011; Varenna, Italy), at the 2011 European School of High Energy Physics (ESHEP 2011; Cheila Gradistei, Romania), and, in modified form, at other summer schools, we present here a written description of the physics of neutrino oscillation. This description is centered on a new way of deriving the oscillation probability. We also provide a brief guide to references relevant to topics other than neutrino oscillation that were covered in the lectures.
\end{abstract}
 
\section{Introduction}
 
If someone asks you what neutrinos are good for, then you can point out that, if neutrinos did not exist, the chain of nuclear reactions that powers the sun would be impossible. The reaction that initiates this chain is the fusion process

$
\begin{array}{ccccccclcl}
          & p &+& p &\ra & d &+ &e^+ & + & \nu  \nonumber  \\
\mathrm{Spin:}  &\frac{1}{2} & &\frac{1}{2}  && 1 & &\frac{1}{2} & &\frac{1}{2}~~~  , \nonumber
\end{array}
$

\vspace{0.2cm}
\hspace{-1.1cm}where we have indicated below each particle its intrinsic spin. Obviously, if a neutrino were not emitted, this process would not conserve angular momentum, so it would be forbidden. Then the chain of reactions that powers the sun could not even get started, and we humans on planet Earth would not exist.

Neutrinos and photons are by far the most abundant elementary particles in the universe. Thus, if we would like to comprehend the universe, we must understand the neutrinos. Of course, studying the neutrinos is challenging, since the only known forces through which these electrically-neutral leptons interact are the weak force and gravity. Consequently, interactions of neutrinos in a detector are very rare events, so that very large detectors and intense neutrino sources are needed to make experiments feasible. Nevertheless, we have confirmed that the weak interactions of neutrinos are correctly described by the Standard Model (SM) of elementary particle physics. Moreover, in the last 13 years, we have discovered that neutrinos have nonzero masses, and that leptons mix. These discoveries have been based on the observation that neutrinos can change from one "flavor" to another --- the phenomenon known as neutrino oscillation. We shall explain the physics of neutrino oscillation, deriving the probability of oscillation in a new way. We shall also provide a very brief guide to references that can be used to study some major neutrino-physics topics other than neutrino oscillation.

\section{Physics of Neutrino Oscillation}
\subsection{Preliminaries}\label{sec2.1}

There are three known flavors of neutrinos: $\nu_e, \nu_\mu$, and $\nu_\tau$. We shall define the neutrino of a given flavor in terms of leptonic $W$-boson decay. This decay produces a charged lepton, which may be an e, $\mu$, or $\tau$, plus a neutrino. We define the $\ne$ as the neutrino produced when the charged lepton is an $e$, the $\nm$ as the neutrino produced together with a $\mu$, and the $\nt$ as the one that accompanies a $\tau$.

Suppose a neutrino $\nu_\alpha$ of flavor $\alpha$ ($= e, \mu,$ or $\tau$), born in a $W$ decay, interacts in a detector {\em immediately,} before it has time to evolve into something else. Suppose further that, by exchanging a $W$ boson with its target in the detector, this neutrino turns into a charged lepton. Then, as far as we know, this charged lepton will always be of the same flavor as the neutrino. Thus, it will be of the same flavor as the charged lepton with which the neutrino was born.

Now imagine that we send a neutrino on a {\em long} journey, say from your present location straight downward to a detector on the opposite side of the Earth. Suppose that this neutrino is created in the pion decay $\pi \ra $ Virtual $W \ra \mu + \nm$, so that at birth it is a $\nm$. Imagine that this neutrino interacts via $W$ exchange in the distant detector, turning into a charged lepton. If neutrinos have masses and leptons mix, then this charged lepton need not be a $\mu$, but could be, say, a $\tau$. Since it is only a $\nt$ that can turn into a $\tau$, the appearance of this $\tau$ would imply that during its journey, our neutrino has evolved from a $\nm$ into a $\nt$, or at least into a neutrino with a nonzero $\nt$ component. The last 13 years have brought us compelling evidence that such changes of neutrino flavor actually occur. As we shall see, the probability of flavor change in vacuum has an oscillatory character, so flavor change is commonly referred to as neutrino oscillation.

That neutrinos have masses means that there is some spectrum of neutrino mass eigenstates $\nu_i$, whose masses $m_i^\nu$ we would like to determine. That leptons mix means that the neutrinos of definite flavor, $\ne, \nm$, and $\nt$, are not the mass eigenstates $\nu_i$. Instead, the neutrino state $\ket{\nu_\alpha}$ of flavor $\alpha$, which is the neutrino state that is created in leptonic $W$ decay together with the charged lepton of the same flavor, is a quantum superposition
\beq
\ket{\nu_\alpha} = \sum_i U^*_{\alpha i} \ket{\nu_i}
\label{eq1}
\eeq
of the mass eigenstates $\ket{\nu_i}$. (From now on, a $\nu$ with a  Greek subscript such as $\alpha$ or $\beta$ will denote a neutrino of definite flavor, while one with a Latin subscript such as $i$ or $j$ will denote a neutrino of definite mass.)
In the superposition of \Eq{1}, the coefficients $U^*_{\alpha i}$ are (complex conjugates of the) elements of the leptonic mixing matrix $U$ --- the leptonic analogue of the quark mixing matrix. Now, there are at least 3 neutrinos $\nu_\alpha$ of definite flavor, and they must be orthogonal to one another, or a neutrino of one flavor, interacting via $W$ exchange, would sometimes turn into a charged lepton of a {\em different} flavor. Out of these 3 orthogonal $\nu_\alpha$, we can form 3 orthogonal linear combinations that will be neutrino mass eigenstates $\nu_i$. (The mass eigenstates must be orthogonal because they are eigenstates of a Hermitean operator, the Hamiltonian.) For all we know, there are more than 3 neutrino mass eigenstates. However, if there are only 3, then $U$ is a 3 $\times$ 3 matrix, and, being the matrix that transforms the states of one quntum basis into those of another, it is unitary. The matrix $U$ is sometimes referred to as the Maki-Nakagawa-Sakata (MNS) matrix, or as the Pontecorvo-Maki-Nakagawa-Sakata (PMNS) matrix, to honor several pioneering contributors to the physics of mixing and oscillation. 

Mixing is readily incorporated into the SM description of the coupling of the leptons to the $W$. For this coupling, we have in the SM Lagrangian density the term
\beq
{\cal L}_{\ell \nu W} = -\frac{g}{\sqrt{2}} \sum_{\alpha = e, \mu, \tau} (\overline{\ell_{L\alpha}} \gamma^\lambda \nu_{L \alpha} W^-_\lambda + \overline{\nu_{L\alpha}} \gamma^\lambda \ell_{L \alpha} W^+_\lambda) ~~.
\label{eq2}
\eeq
Here, $g$ is the semi-weak coupling constant, $\nu_\alpha$ is the neutrino of flavor $\alpha$ as before, and $\ell_\alpha$ is the charged lepton of flavor $\alpha$. That is, $\ell_e \equiv e, \;  \ell_\mu \equiv \mu$, and $\ell_\tau \equiv \tau$. The subscript $L$ denotes a left-handed chiral projection: $\ell_{L\alpha} = [(1-\gamma_5)/2] \ell_\alpha$, and similarly for $\nu_{L\alpha}$. Note from \Eq{2} that, in conformity with the rule quoted earlier, the neutrino of flavor $\alpha$ couples only to the charged lepton of the same flavor. To explicitly incorporate mixing into the $\ell\,\nu\,W$ coupling, we insert \Eq{1} into \Eq{2}, so that the latter becomes
\beq
{\cal L}_{\ell \nu W} = -\frac{g}{\sqrt{2}} 
\sum_{\begin{array}{c} {\st \alpha = e, \mu, \tau} \vspace{-0.05in} \\  {\st i = 1, 2, 3}  \end{array} }
(\overline{\ell_{L\alpha}} \gamma^\lambda U_{\alpha i} \nu_{L i} W^-_\lambda + \overline{\nu_{Li}} \gamma^\lambda U^*_{\alpha i} \ell_{L \alpha} W^+_\lambda) ~~.
\label{eq3}
\eeq
Here, as before, $\nu_i$ is a neutrino mass eigenstate, and we have taken into account the fact that the field operator which absorbs the state $\sum_j U^*_{\alpha j} \ket{\nu_j}$ of \Eq{1} is not $\sum_i U^*_{\alpha i} \nu_i$, but $\sum_i U_{\alpha i} \nu_i$.

From \Eq{3}, we see that, apart from the factor of $g/\sqrt{2}$ out front and kinematical factors, the amplitude for $W^+ \ra \ell^+_\alpha + \nu_i$ or for $\ell^-_\alpha + W^+ \ra \nu_i$ is just $U^*_{\alpha i}$, while that for $\nu_i \ra \ell^-_\beta + W^+$ is just $U_{\beta i}$. Writing out the mixing matrix explicitly, it is
\begin{eqnarray}  
    &    & {}\hspace{1.07cm}  \nu_1  \hspace{0.6cm}  \nu_2  \hspace{0.6cm} \nu_3  \nonumber    \\
U & = & \begin{array}{c}  e \\ \mu \\ \tau \end{array}     \hspace{-0.05in}
	 \left[   \begin{array}{ccc}
	U_{e1}      & U_{e2}       & U_{e3}        \\
	U_{\mu 1} & U_{\mu 2} & U_{\mu 3}    \\
	U_{\tau 1} & U_{\tau 2} & U_{\tau 3}
	\end{array}  \right]	~~ .
\label{eq4}
\end{eqnarray}

From \Eq{3}, we see that, for example, the $e$ (top) row of $U$ tells us what linear combination of the neutrino mass eigenstates $\nu_1,\; \nu_2$, and $\nu_3$ couples to an $e$ and a $W$. Similarly, the $\nu_1$ (first) column of $U$ tells us what linear combination of the charged lepton mass eigenstates $e,\; \mu$, and $\tau$ couples to a $\nu_1$ and a $W$. Similarly for the other rows and columns.

\subsection{Probability of Neutrino Oscillation in Vacuum}\label{sec2.2}

\begin{figure}[htb]
\centering
\includegraphics[width=.9\linewidth]{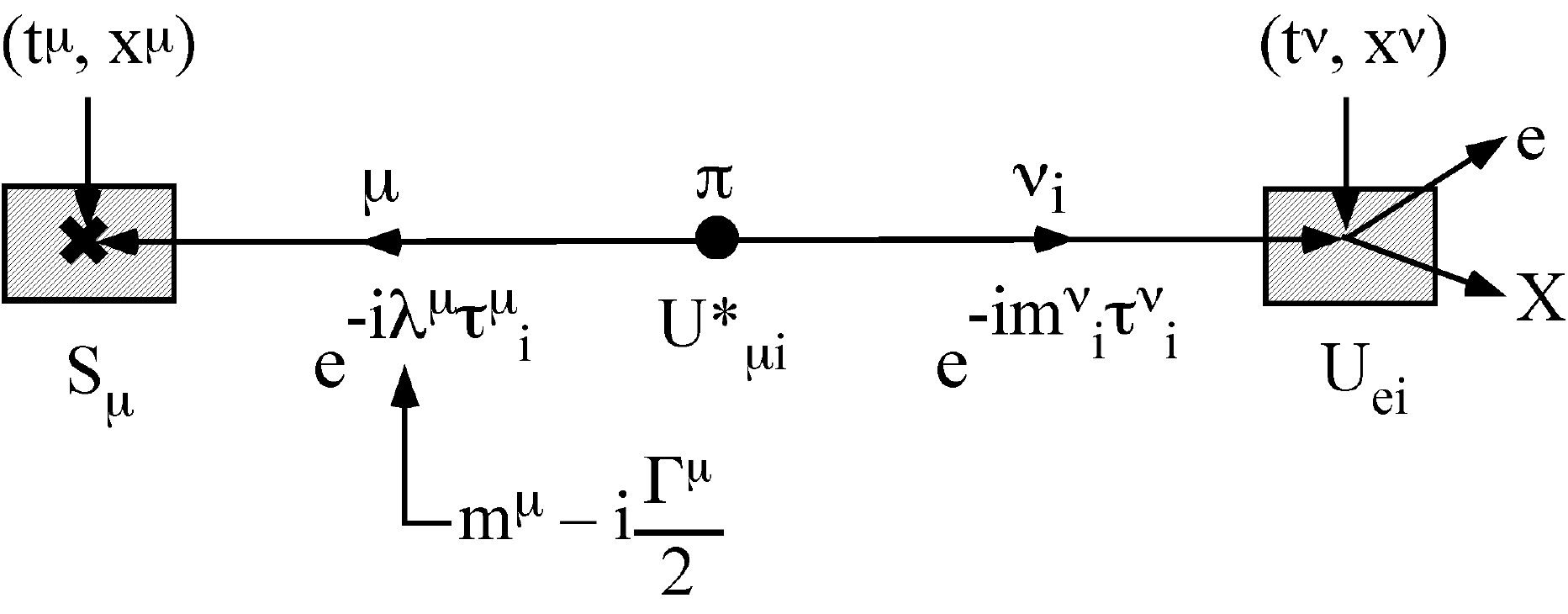}
\caption{The pion-rest-frame view of a pion decay followed by interactions of the daughter neutrino and muon. The amplitudes shown for the various parts of this scenario are explained in the text.}
\label{f1}
\end{figure}

Let us now find the probability P($\nu_\alpha \ra \nu_\beta; \:L,E$) that a neutrino born as a $\nu_\alpha$ --- a neutrino of flavor $\alpha$ --- will then behave like a $\nu_\beta$ --- a neutrino of flavor $\beta$ --- after traveling through vacuum for a distance $L$ with energy $E$. The conventional derivation of this probability may be found in many places in the literature \cite{ref1}. Rather than reproduce that derivation, or present the one we gave in the lectures (which may be found in Ref. \cite{ref2}), here we present a new approach \cite{ref3}. We apply this approach to neutrinos produced in the pion decay $\pi \ra \mu + \nu$, which we view in the pion rest frame, as in Fig.\ ~\ref{f1}.
To illustrate how the approach works, we consider a scenario in which the neutrino, having been created in the $\pi$ decay at a spacetime point $(0,\,0)$, then interacts at the spacetime point $(t^\nu,\, x^\nu)$ whose pion-rest-frame coordinates are $t^\nu$ and $x^\nu$. The interaction is via $W$ exchange with a target in a neutrino detector, and, for illustration, we suppose that the charged lepton into which the neutrino is converted by this interaction is an electron. Of course, the interaction will also produce a recoil $X$. 
As part of the same overall scenario, the muon created together with the neutrino in the $\pi$ decay interacts at the spacetime point $(t^\mu,\, x^\mu)$ whose pion-rest-frame coordinates are $t^\mu$ and $x^\mu$. We imagine that the interaction is with matter that surrounds the $\pi$ decay region. The full scenario is pictured in Fig.\ ~\ref{f1}.

We shall find the amplitude for the entire scenario, including the $\pi$ decay and the $\nu$ and $\mu$ interactions \cite{ref4}. To do this, we shall use the fact that if a particle has mass $m$ and width $\Gamma$, adding up to a complex mass $\lambda = m - i\Gamma /2$, then the amplitude for this particle to propagate for a proper time $\tau$ in its own rest frame is $\exp(-i \lambda \tau)$. (If this propagation is through a distance $x$ during a time $t$ in some frame in which the particle has momentum $p$ and energy $E$, then $\exp (-im\tau)$, the non-decaying part of $\exp (-i\lambda\tau)$, is simply the familiar quantum-mechanical plane-wave factor $\exp [i(px-Et)])$.

In the neutrino mass eigenstate basis, the neutrino that travels from the $\pi$ decay point to the interaction point in the neutrino detector will be one or another of the neutrino mass eigenstates $\nu_i$. Since we cannot make measurements that determine which $\nu_i$ was involved in any given event without destroying the oscillation pattern, we must add the amplitudes for the contributions of the different $\nu_i$ coherently.

From \Eq{3} and the discussion that follows it, the relevant factor in the amplitude for the decay $\pi \ra \mu + \nu$ to yield, in particular, the neutrino mass eigenstate $\nu_i$ is $U^*_{\mu i}$, as shown in Fig.\ ~\ref{f1}. Similarly, the relevant factor for the charged lepton created when the $\ni$ interacts in the detector to be, in particular, an $e$ is $U_{ei}$. The amplitude for the muon to interact in matter we shall call $S_\mu$.

From the discussion of the amplitude for a particle to propagate, we see that the amplitude for the neutrino mass eigenstate $\nu_i$, of mass $m^\nu_i$, to propagate is $\exp(-im_i^\nu \tau_i^\nu )$. Here, we are neglecting the extremely small neutrino decay width, and $\tau_i^\nu$ is the proper time that elapses in the $\nu_i$ rest frame while the $\nu_i$ travels from the spacetime point $(0,\, 0)$ where it was born to the given interaction point $(t^\nu,\, x^\nu)$ in the detector. 
The index $i$ on $\tau_i^\nu$ is present because the time that elapses in the $\nu_i$ rest frame during the journey to the given point $(t^\nu,\, x^\nu)$ in the pion rest frame depends on the $\ni$ energy in the latter frame, hence on the $\ni$ mass, and consequently on which $\ni$ is involved.

Similarly, the amplitude for the muon to propagate is $\exp(-i\lambda^\mu \tau_i^\mu) = \exp [-i(m^\mu -i\Gamma^\mu /2)\tau_i^\mu]$. Here we are taking into account the decay of the muon by including in its complex mass $\lambda^\mu$ its decay width $\Gamma^\mu$. 
The quantity $\tau_i^\mu$ is the proper time that elapses in the $\mu$ rest frame while the $\mu$ travels from the $\pi$ decay point $(0,\, 0)$ to the $\mu$ interaction point $(t^\mu,\, x^\mu)$. The index $i$ on $\tau_i^\mu$ is present because the muon and the neutrino are kinematically entangled. The energies of both of these particles in the pion rest frame depend on the mass of the emitted neutrino, so that they depend on which $\ni$ it is. Consequently, the proper times that elapse in the rest frames of the muon and the neutrino both depend on which $\nu_i$ is emitted, at least in principle.

Multiplying together the amplitudes for the various parts of the scenario pictured in Fig.\ ~\ref{f1}, and coherently adding the contributions of the different $\ni$, we find that the amplitude {\it Amp} for the entire scenario is given by
\beq
\mathit{Amp} = \sum_{i=1,2,3} S_\mu\, e^{-i(m^\mu -i\Gamma^\mu /2)\tau_i^\mu)} U^*_{\mu i}\, e^{-im_i^\nu \tau_i^\nu} U_{ei} ~~.
\label{eq5}
\eeq
We note that this amplitude is Lorentz invariant.

How do the muon and neutrino propagation amplitudes $\exp[-i(m^\mu -i\Gamma^\mu /2)\tau_i^\mu]$ and $\exp(-im_i^\nu \tau_i^\nu)$ actually depend on $i$? From the Lorentz transformation, the proper time $\tau_i^\mu$ in the muon propagation amplitude is given by 
\beq
\tau_i^\mu = \frac{1}{m^\mu} (E_i^\mu t^\mu - p_i^\mu x^\mu) ~~.
\label{eq6}
\eeq
Here, $E_i^\mu$ and $p_i^\mu$ are, respectively, the muon energy and momentum in the pion rest frame when the emitted neutrino is $\ni$. In evaluating the right-hand side of \Eq{6}, we choose a muon interaction point at which $x^\mu$ is related to $t^\mu$ by
\beq
x^\mu = v_0^\mu t^\mu  = \frac{p_0^\mu}{E_0^\mu} t^\mu ~~.
\label{eq7}
\eeq
Here, $v_0^\mu,\, p_0^\mu$, and $E_0^\mu$ are, respectively, the velocity, momentum, and energy that the muon would have in the pion rest frame if neutrinos were massless. The spacetime points passed by the peak of the quantum-mechanical wave packet that describes the muon propagation in greater detail than is needed here would satisfy \Eq{7} to an excellent approximation.

From Eqs. (\ref{eq6}) and (\ref{eq7}), the muon in $\pi \ra \mu + \ni$ and that in $\pi\ra\mu + \nu_j$ have travel proper times that differ by
\beq
\tau_i^\mu - \tau_j^\mu = \frac{t^\mu}{m^\mu} \left[ (E^\mu_i - E^\mu_j) - (p^\mu_i - p^\mu_j) \frac{p^\mu_0}{E^\mu_0} \right] ~~.
\label{eq8}
\eeq
For a given neutrino mass $m^\nu$, the pion-rest-frame energy of the muon in $\pi\ra\mu+\nu$ is
\beq
E^\mu = \frac{ (m^\pi)^2 + (m^\mu)^2 - (m^\nu)^2}{2 m^\pi} ~~,
\label{eq9}
\eeq
where $m^\pi$ is the pion mass. Thus, in \Eq{8}, 
\beq
E^\mu_i - E^\mu_j = -\frac{\Delta m^2_{ij}}{2m^\pi} ~~,
\label{eq10}
\eeq
where $\Delta m^2_{ij} \equiv (m^\nu_i)^2 -  (m^\nu_j)^2$. Moreover, for given muon energy $E^\mu,\; (p^\mu)^2 = (E^\mu)^2 - (m^\mu)^2$, so that
\beq
\frac{dp^\mu}{d[(m^\nu)^2]} = \frac{E^\mu}{p^\mu} \, \frac{dE^\mu}{d[(m^\nu)^2] } ~~ .
\label{eq11}
\eeq
Consequently, to lowest order in the squares of the neutrino masses,
\beq
p^\mu_i - p^\mu_j = \frac{E^\mu_0}{p^\mu_0} \left[ - \frac{\Delta m^2_{ij}}{2m^\pi} \right]~~.
\label{eq12}
\eeq
Inserting Eqs. (\ref{eq10}) and (\ref{eq12}) into \Eq{8}, we find that to lowest (i.e. first) order in the squares of the neutrino masses,
\beq
\tau_i^\mu - \tau_j^\mu = \frac{t^\mu}{m^\mu} \left[  -\frac{\Delta m^2_{ij}}{2m^\pi}\right] \left[ 1 -   \frac{E^\mu_0}{p^\mu_0}\frac{p^\mu_0}{E^\mu_0} \right]  = 0  ~~.
\label{eq13}
\eeq
Thus, to lowest order, the muon propagation amplitude $\exp(-im_i^\mu \tau_i^\mu) \exp [-(\Gamma^\mu /2) \tau^\mu_i]$ actually does not depend on which $\ni$ is emitted \cite{ref5}. The factor $\exp(-im_i^\mu \tau_i^\mu)$ will have no significant effect at all on the absolute square of the amplitude of \Eq{5} for the scenario in Fig.\ \ref{f1}.
The factor $ \exp [-(\Gamma^\mu /2) \tau^\mu_i]$ will lead to an overall decay of this amplitude with muon travel time, reflecting the obvious fact that the probability for the muon to remain present (i.e., not yet decayed), so that it may interact, decays with time. But this overall decay of the amplitude for the scenario in Fig.\ ~\ref{f1} will not affect the neutrino oscillation pattern. Since it is only that pattern in which we are ultimately interested, we can drop the entire muon propagation amplitude from the amplitude of \Eq{5}.

Turning to the propagation amplitude $\exp(-im_i^\nu \tau_i^\nu)$ for the neutrino $\ni$, we have from the Lorentz transformation the relation
\beq
m_i^\nu \tau_i^\nu = E^\nu_i t^\nu - p^\nu_i x^\nu ~~.
\label{eq14}
\eeq
Here, $E^\nu_i$ and $p^\nu_i$ are, respectively, the $\ni$ energy and momentum in the pion rest frame. Since, in practice, neutrinos are ultra-relativistic, we choose a neutrino interaction point at which $t^\nu$ is related to $x^\nu$ by $t^\nu = x^\nu \equiv L^0$. Then the propagation phases for neutrino mass eigenstates $\ni$ and $\nu_j$ differ by
\beq
m_i^\nu \tau_i^\nu - m_j^\nu \tau_j^\nu = [ (E^\nu_i - E^\nu_j) -(p^\nu_i-p^\nu_j)] L^0 ~~.
\label{eq15}
\eeq
In analogy to \Eq{9}, for given neutrino mass $m^\nu$, the pion-rest-frame energy of the neutrino, $E^\nu$, is given by
\beq
E^\nu = \frac{ (m^\pi)^2 + (m^\nu)^2 - (m^\mu)^2}{2 m^\pi} ~~.
\label{eq16}
\eeq
Thus, the energies of two different neutrino mass eigenstates $\ni$ and $\nu_j$ differ by
\beq
E^\nu_i - E^\nu_j = \frac{\Delta m^2_{ij}}{2m^\pi} ~~.
\label{eq17}
\eeq
In addition, for given neutrino energy $E^\nu,\; (p^\nu)^2 = (E^\nu)^2 - (m^\nu)^2$. From this relation and conservation of energy in $\pi\ra\mu+\nu$, one easily finds that
\beq	
\left. \frac{dp^\nu}{d[m^\nu)^2]} \right|_{m^\nu = 0}= - \frac{E^\mu_0}{E^\nu_0} \frac {1}{2m^\pi} ~~,
\label{eq18}
\eeq
where $E^\nu_0 =  [(m^\pi)^2- (m^\mu)^2] / 2m^\pi$ is the pion-rest-frame energy that the neutrino would have if it were massless. It follows that, to lowest order in $\Delta m^2_{ij}$, the momenta of $\ni$ and $\nu_j$ differ by
\beq
p^\nu_i - p^\nu_j = - \frac{E^\mu_0}{E^\nu_0}\,\frac{\Delta m^2_{ij}}{2m^\pi} ~~.
\label{eq19}
\eeq
Inserting  Eqs. (\ref{eq17}) and (\ref{eq19}) into \Eq{15}, we find that to lowest order,
\beq
m^\nu_i\tau_i^\mu - m^\nu_j\tau_j^\nu = \frac{\Delta m^2_{ij}}{2m^\pi} \left[ 1 +  \frac{E^\mu_0}{E^\nu_0} \right]  L^0 = \Delta m^2_{ij} \frac{L^0}{2E^\nu_0}  ~~.
\label{eq20}
\eeq
From this result, we see that we may take the neutrino propagation amplitude $\exp (-im^\nu_i\tau_i^\nu)$ to be
\beq
e^{-i(m^\nu_i)^2 \frac{L^0}{2E^\nu_0}}  ~~,
\label{eq21}
\eeq
and all the relative phases in the amplitude of \Eq{5} will be correct. Then, if we delete from \Eq{5} the muon interaction and propagation amplitudes, which do not affect the neutrino oscillation pattern because they are $i$-independent, \Eq{5} yields
\beq
\mathit{Amp} = \sum_{i=1,2,3} U^*_{\mu i} e^{-i(m_i^\nu)^2 \frac{L^0}{2E^\nu_0}} U_{ei} ~~.
\label{eq22}
\eeq

A neutrino flavor-change experiment will carry out its work in the rest frame of some neutrino detector --- the laboratory frame. However, until now we have been viewing our illustrative process of interest from the rest frame of the pion whose decay creates our neutrino. Now, as we shall see momentarily, for given neutrino energy, the probability of flavor change oscillates as a function of the distance $L$ that the neutrino travels in the laboratory frame. 
If we are to observe this oscillation, then obviously the neutrino source --- the pion in our example --- must be spacially localized to within an oscillation wavelength. But then, by the uncertainty principle $\Delta p \Delta x \geq \hbar$, there must be some uncertainty in the lab-frame pion momentum \cite{ref6}. The pions whose decays produce the neutrinos of an oscillation experiment cannot be known to be precisely at rest. Thus, we must find the amplitude for the scenario in Fig.\ ~\ref{f1} when the pion is moving in the lab frame --- the rest frame of the neutrino detector. In addition, we must express this amplitude in terms of lab-frame variables. 
Accomplishing these goals is easy. First, we recall that the amplitude {\it Amp} of Eqs. (\ref{eq5}) and (\ref{eq22}) is Lorenz invariant. It is valid both in the pion rest frame and in the lab frame, in which in general the pion is moving. Secondly, to express the amplitude of \Eq{22} in terms of lab-frame, rather than pion-rest-frame, variables, we note that, as already remarked, neutrinos are ultra-relativistic. Thus, in the pion rest frame, the travel time of one of them is equal to its travel distance $L^0$. Then, by the Lorentz transformation, its travel distance $L$ in the lab frame is given by
\beq
L = \gamma_\pi (1 + \beta_\pi) L^0 ~~,
\label{eq23}
\eeq
where $\beta_\pi$ is the velocity of the pion in the lab, and $\gamma_\pi = 1 / \sqrt{1 - \beta_\pi ^2}$. Similarly, in the pion rest frame, the momentum $p_0^\nu$ of a massless neutrino is equal to its energy $E_0^\nu$. Thus, its energy $E$ in the lab frame is given by 
\beq
E = \gamma_\pi (1 + \beta_\pi) E_0^\nu ~~.
\label{eq24}
\eeq
We see that
\beq
\frac{L}{E} = \frac{L^0}{E_0^\nu} ~~,
\label{eq25}
\eeq
so that we may write the amplitude of \Eq{22} as
\beq
\mathit{Amp} = \sum_{i=1,2,3} U^*_{\mu i}\, e^{-i(m_i^\nu)^2 \frac{L}{2E}} \,U_{ei} ~~.
\label{eq26}
\eeq

As explained in Sec. \ref{sec2.1}, the neutrino produced in $\pi\ra\mu+\nu$ is by definition a $\nu_\mu$. In the calculation above, we have worked in neutrino mass eigenstate basis, so, in effect, we have broken the $\nm$ down into its mass eigenstate components. For purposes of illustration, we have considered a scenario in which the neutrino interaction in the detector yields an electron. Since, in neutrino flavor basis, it is only a $\ne$ that can yield an electron, the sequence of events pictured in Fig.\ ~\ref{f1} is what would commonly be called $\nm \ra \ne$ oscillation, with the addition of an interaction between matter and the muon that is produced together with the neutrino in the pion decay. 
We are interested mainly in the probability for the $\nm \ra \ne$ oscillation, integrated over all the possible fates of the muon. Apart from a possible overall normalization factor, this muon-integrated $\nm \ra \ne$ oscillation probability, P$(\nm \ra \ne; \, L, E)$, will be given by the absolute square of the amplitude {\it Amp} of \Eq{26}, from which the muon interaction and propagation amplitudes have been removed. 

Generalizing to a scenario in which the neutrino is born together with a charged lepton of flavor $\alpha\; (=e,\, \mu$, or $\tau)$, and then interacts in a detector and makes a charged lepton of flavor $\beta$ (not necessarily different from $\alpha$), we see from \Eq{26} that the amplitude would be
\beq
\mathit{Amp}(\nu_\alpha \ra \nu_\beta; \, L, E) = \sum_{i=1,2,3} U^*_{\alpha i} \, e^{-i(m_i^\nu)^2 \frac{L}{2E}} U_{\beta i} ~~.
\label{eq27}
\eeq
Apart from a possible overall normalization factor, the probability P$(\nu_\alpha \ra \nu_\beta; \, L, E)$ of the $\nu_\alpha \ra \nu_\beta$ oscillation will then be the absolute square of this amplitude. Assuming that the mixing matrix $U$ is unitary, we find from \Eq{27} that this absolute square, summed over all possible final flavors $\beta$, including $\beta = \alpha$, is
\begin{eqnarray}
\sum_{\mathrm{All}\; \beta} | \mathit{Amp}(\nu_\alpha \ra \nu_\beta; \, L, E)|^2 & = &  \sum_\beta \left( \sum_i U^*_{\alpha i} e^{-i(m_i^\nu)^2 \frac{L}{2E}} U_{\beta i}\right)
\left( \sum_jU_{\alpha j} e^{i(m_j^\nu)^2 \frac{L}{2E}} U^*_{\beta j}\right)  \nonumber \\
	& = & \sum_{i,j} U^*_{\alpha i}\, e^{-i(m_i^\nu)^2 \frac{L}{2E}}\, U_{\alpha j}\,
		 e^{i(m_j^\nu)^2 \frac{L}{2E}} \delta_{ij}     \nonumber \\
	& = & \sum_i |U_{\alpha i}|^2 = 1 ~~.
\label{eq28}
\end{eqnarray}
Thus, the amplitude {\it Amp}$(\nu_\alpha \ra \nu_\beta; \, L, E)$ of \Eq{27} is a properly normalized probability amplitude. It needs no additional normalization factor. The probability P$(\nu_\alpha \ra \nu_\beta; \, L, E)$ of $\nu_\alpha \ra \nu_\beta$ oscillation is simply its absolute square. Taking this absolute square, and making use of the assumed unitarity of the mixing matrix, we find that
\begin{eqnarray}
\mathrm{P}(\nu_\alpha \ra \nu_\beta; \, L, E) & = & \delta_{\alpha \beta} - 4\sum_{i>j}
	\real (U^*_{\alpha i} U_{\beta i} U_{\alpha j} U^*_{\beta j}) \sin^2 ( \Delta m^2_{ij} \frac{L}{4E})      \nonumber \\
	&& \phantom{\delta_{\alpha \beta}} + 2\sum_{i>j}
	\imag (U^*_{\alpha i} U_{\beta i} U_{\alpha j} U^*_{\beta j}) \sin ( \Delta m^2_{ij} \frac{L}{2E}) ~~.
\label{eq29}
\end{eqnarray}

In deriving this expression for the oscillation probability, we have assumed that the neutral lepton in Fig.~\ref{f1} is a {\em neutrino}, not an {\em antineutrino.} The factors $U^*_{\mu i}$ and $U_{ei}$ that we took from \Eq{3} and incorporated into the amplitude of \Eq{5} depended on this assumption. As we see from \Eq{3}, if the neutral lepton had been an antineutrino, then $U^*_{\mu i}$ and $U_{ei}$ would have been replaced, respectively, by $U_{\mu i}$ and $U^*_{ei}$. 
In addition, the amplitude $S_\mu$ for the matter interaction of the $\mu^+$ from the reaction $\pi^+ \ra \mu^+ + \nu$ that produces a neutrino would have been replaced by a different amplitude $S_\mu ^\prime$ for the matter interaction of the $\mu^-$ from the reaction $\pi^- \ra \mu^- + \bar{\nu}$ that produces an antineutrino. 
However, as we have seen, the muon-matter interaction amplitude is ultimately irrelevant. Moreover, so long as CPT invariance holds, a particle and its antiparticle have the same mass and the same width. Thus, the muon and neutrino propagation amplitudes in \Eq{5} would be unchanged if the $\mu^+$ and neutrino from $\pi^+ \ra \mu^+ + \nu$ were replaced by the $\mu^-$ and antineutrino from $\pi^- \ra \mu^- + \bar{\nu}$. We conclude from this $\pi \ra \mu + \nu$ example that, completely generally, 
\beq
\mathrm{P}(\overline{\nu_\alpha} \ra \overline{\nu_\beta}; \, L, E) = \mathrm{P}(\nu_\alpha \ra \nu_\beta; \, L, E\: | \:U \ra U^*)  ~~.
\label{eq30}
\eeq
That is, the probability for the antineutrino oscillation $\overline{\nu_\alpha} \ra \overline{\nu_\beta}$ is the same as for the corresponding neutrino oscillation $\nu_\alpha \ra \nu_\beta$, except that the mixing matrix $U$ in the latter is replaced by $U^*$ in the former. From \Eq{29}, we then have 
\begin{eqnarray}
\mathrm{P}\optbar{\nu_\alpha} \ra \optbar{\nu_\beta}; \, L, E) & = & \delta_{\alpha \beta} - 4\sum_{i>j}
	\real (U^*_{\alpha i} U_{\beta i} U_{\alpha j} U^*_{\beta j}) \sin^2 ( \Delta m^2_{ij} \frac{L}{4E})      \nonumber \\
	&& \phantom{\delta_{\alpha \beta}} \pom 2\sum_{i>j}
	\imag (U^*_{\alpha i} U_{\beta i} U_{\alpha j} U^*_{\beta j}) \sin ( \Delta m^2_{ij} \frac{L}{2E}) ~~.
\label{eq31}
\end{eqnarray}
We see that if $U$ is not real, then the $\nu_\alpha \ra \nu_\beta$ and $\overline{\nu_\alpha} \ra \overline{\nu_\beta}$ oscillations probabilities can differ.

The coupling of the leptons to the $W$ boson has a parity-violating, chirally left-handed structure, as described by \Eq{3} or \Eq{2}. Moreover, the neutrinos we study experimentally are ultra-relativistic, and for ultra-relativistic fermions there is essentially no difference between chirality and helicity. As a result, the neutrinos we study experimentally, which are produced by the chirally left-handed coupling of \Eq{3}, are essentially always of left-handed (i.e., negative) helicity. It is easy to show that, in contrast, the antineutrinos produced by this coupling are essentially always of right-handed (i.e., positive) helicity. By  $\nu_\alpha \ra \nu_\beta$, we mean the oscillation of neutrinos of left-handed helicity, and by $\overline{\nu_\alpha} \ra \overline{\nu_\beta}$ we mean the oscillation of antineutrinos of right-handed helicity. 
Now, the particle-antiparticle symmetry operation that turns a neutrino of left-handed helicity into an antineutrino of right-handed helicity is CP. The charge conjugation operation C turns the neutrino into an antineutrino with no change of kinematical variables, and the parity operation P reverses the helicity. Thus, if, owing to a nonvanishing value of the last term in \Eq{31}, the probabilities for  $\nu_\alpha \ra \nu_\beta$ and $\overline{\nu_\alpha} \ra \overline{\nu_\beta}$ should differ, this difference would be a violation of CP invariance. To date, CP violation has been observed only in the quark sector, and its observation in the neutrino sector would establish that the leptons violate CP as well. 
This observation in the latter sector would also make it more plausible that the baryon-antibaryon asymmetry of the universe arose, at least in part, through a scenario called leptogenesis that involves hypothesized very heavy neutrinos \cite{ref7, ref8}.

We see from \Eq{31} that the oscillation probability oscillates as a function of $L/E$, justifying our calling neutrino flavor change ``oscillation''. We also see from \Eq{31} that oscillation from one flavor $\alpha$ into a different one $\beta$ implies nonzero mass splittings $\Delta m^2_{ij}$, hence nonzero neutrino masses. Similarly, such oscillation implies that $U$ is not diagonal, which is to say that there is nontrivial leptonic mixing. Inserting into the quantity $\Delta m^2_{ij}\, L/4E$, on which the oscillation in \Eq{31} depends, the so-far-omitted factors of $\hbar$ and $c$, we find that 
\beq
\Delta m^2_{ij} \frac{L}{4E} = 1.27 \Delta m^2_{ij} (\mathrm{eV}^2) \frac{L \:\mathrm{(km)}}{E \:\mathrm{(GeV)}} ~~.
\label{eq32}
\eeq
Thus, the factor $\sin^2 ( \Delta m^2_{ij} \frac{L}{4E})$ in \Eq{31} becomes $\sin^2 [1.27 \Delta m^2_{ij} (\mathrm{eV}^2) \frac{L \:\mathrm{(km)}}{E \:\mathrm{(GeV)}}]$. This factor is appreciable when its argument is $\geq {\cal O} (1)$. Thus, an oscillation experiment with given $L/E$ is sensitive to squared-mass splittings $\Delta m^2_{ij} (\mathrm{eV}^2) \gsim E \,\mathrm{(GeV)} / L \,\mathrm{(km)}$. For example, if $E = 1$ GeV, and $L$ is the diameter of the Earth, $\sim 10^4$ km, values of $E$ and $L$ that are encountered in studies of the neutrinos made in the Earth's atmosphere by cosmic rays, then there will be sensitivity to $\Delta m^2_{ij} \gsim 10^{-4}$ eV$^2$.
As this illustrates, neutrino oscillation experiments can be sensitive to very tiny mass splittings. We note, however, that oscillation depends {\em only} on these splittings, and not on the individual neutrino masses. Determining those individual masses will require another approach.

Neutrino flavor change can be sought in two ways. In a beam of neutrinos born with a known flavor $\alpha$, one can look for the appearance of neutrinos of a different flavor $\beta$. This is referred to as an appearance experiment. Alternatively, in a known flux of neutrinos $\nu_\alpha$ of a given flavor $\alpha$, one can look for the disappearance of some of this known flux, due to the oscillation of some of the $\nu_\alpha$ into neutrinos of other flavors. This is referred to as a disappearance experiment.

In \Eq{28}, we have confirmed that the probability of oscillation, $\mathrm{P}(\nu_\alpha \ra \nu_\beta; \, L, E)$, summed over all possible final flavors $\beta$, including $\beta=\alpha$, is unity. That is, the probability that a neutrino changes flavor plus the probability that it does not change flavor is unity. 
This statement remains true even if there are more than the three flavors of neutrino $(\ne,\; \nm$, and $\nt)$ that we have been taking into account. However, from experimental studies of the decays $Z \ra \nu_\alpha \, \overline{\nu_\alpha}$ of the $Z$ boson, we know that any additional flavors of neutrino beyond $\ne,\; \nm$, and $\nt$ do not couple to the $Z$. The SM then implies that these additional flavors do not couple to the $W$ either. Neutrinos that do not couple to the SM $W$ or $Z$ bosons, and consequently do not participate in any known interaction other than gravity, are called sterile neutrinos. Such neutrinos may participate in some as-yet-unknown interaction that lies beyond the SM. However, any such interaction is invisible at presently accessible neutrino energies, so sterile neutrinos will leave no trace in a neutrino detector. 
Thus, if we start with a beam of neutrinos of one of the active flavors (i.e., $\ne,\; \nm$, or $\nt)$, and some of the neutrinos in this beam oscillate into sterile neutrinos, an experiment that can measure the {\em total} active flux in the beam (i.e., the sum of the $\ne,\; \nm$, and $\nt$ fluxes) will find that some of the active flux has vanished.

Among the special cases of the oscillation probability formula of \Eq{31}, the best known is the one that describes oscillation when only two mass eigenstates are important. Let us call these mass eigenstates $\nu_1$ and $\nu_2$, and the two neutrinos of definite flavor that we can construct as superpositions of $\nu_1$ and $\nu_2$, $\nu_\alpha$ and $\nu_\beta$. There is only one squared-mass splitting, $m^2_2 - m^2_1 \equiv \Delta m^2$, in this physical system, and the mixing matrix $U$ is 2 $\times$ 2. It can be shown that, as far as neutrino oscillation is concerned, if $U$ is unitary, it may be taken to be given by
\beq
U \equiv \left[  \begin{array}{cc}
	U_{\alpha  1} & U_{\alpha 2}  \\  U_{\beta  1} & U_{\beta 2}  \end{array} \right]
	= \left[  \begin{array}{cc}
	\phantom{-} \cos\theta & \sin\theta  \\  -\sin\theta & \cos\theta   \end{array} \right] ~~.
\label{eq33}
\eeq
In this expression, the angle $\theta$ is referred to as the mixing angle. Inserting this mixing matrix and the single $\Delta m^2$ into \Eq{31}, we find immediately that for $\beta \neq \alpha$
\beq
\mathrm{P} (\optbar{\nu_\alpha} \ra \optbar{\nu_\beta} ) = \sin^2 2\theta \, \sin^2 (\Delta m^2 \frac{L}{4E}) ~~.
\label{eq34}
\eeq
For no flavor change, we find that
\beq
\mathrm{P} (\optbar{\nu_\alpha} \ra \optbar{\nu_\alpha} ) = 1 - \sin^2 2\theta \, \sin^2 (\Delta m^2 \frac{L}{4E}) ~~.
\label{eq35}
\eeq

\subsection{Neutrino Flavor Change in Matter}\label{sec2.3}

Many of the experimental studies of neutrino flavor change that have been carried out have involved neutrinos that travel through matter. In some cases, interaction of the neutrinos with electrons in the matter significantly modifies the flavor content of the beam, relative to what it would be in vacuum. Treatment of the effect of matter on neutrino flavor change may be found, for example, in Refs. \cite{ref1} and \cite{ref9}. Here we shall make just a few brief comments.

Coherent forward scattering of an electron neutrino $\ne$ from electrons in matter, caused by $W$-boson exchange, gives the $\ne$ an extra interaction potential energy
\beq
V = + \sqrt{2} G_F N_e ~~.
\label {eq36}
\eeq
Here, $G_F$ is the Fermi coupling constant of the weak interaction, and $N_e$ is the number of electrons per unit volume. Correspondingly, an electron antineutrino $\overline{\nu_e}$ traveling through matter has an extra interaction potential energy
\beq
\bar{V} = - \sqrt{2} G_F N_e ~~.
\label {eq37}
\eeq
These extra energies raise the effective mass of a $\ne$ in matter, and lower that of a $\overline{\ne}$. A useful measure of the fractional importance of this matter effect on an oscillation involving a vacuum mass splitting $\Delta m^2$ is given by the parameter
\beq
x = \frac{\sqrt{2} G_F N_e}{\Delta m^2 / 2E} ~~.
\label{eq38}
\eeq
On the right hand side of this relation, the numerator is the extra energy of a $\ne$ due to matter interaction, and the denominator is the quantity with dimensions of energy that occurs in the relative phase of two interfering terms in the vacuum oscillation amplitude of \Eq{27}.

We see from \Eq{38} that the matter effect grows with the neutrino energy $E$. As \Eq{38} suggests, the matter effect is sensitive to the sign of $\Delta m^2$. That is, it can be used to determine which of two mass eigenstates with known couplings to the various charged leptons is the heavier one. Owing to the fact that $\bar{V} = -V$, matter affects antineutrinos differently than it affects neutrinos. As a result, an observed difference between the oscillation in matter of antineutrinos and neutrinos can have two sources: 1) CP violation coming from a mixing matrix $U$ that is not real, as may be seen from  \Eq{31}, and 2) the matter effect. Experiments seeking to demonstrate that neutrino oscillation violates CP will have to disentangle these two effects.

\section{A Brief Guide to References}

The lectures presented at the 2011 European School of High Energy Physics, and the 2011 International School on Astro Particle Physics devoted to Neutrino Physics and Astrophysics, covered a number of topics in addition to the physics of neutrino oscillation. One may study those other topics in the following references.

History of neutrino oscillation results: Refs.~\cite{ref1} and \cite{ref10}.

Physics of Majorana neutrinos and neutrinoless double beta decay:  Refs.~\cite{ref11} - \cite{ref13}.

Leptogenesis: Refs.~\cite{ref7, ref8}.

Recent experimental results, and the status of our knowledge: The original papers in this fast-moving field, and Ref.~\cite{ref14}.

\section*{Acknowledgments}

The author appreciates partial support from the European Union FP7 ITN INVISIBLES (Marie Curie Actions, PITN- GA-2011- 289442).


\begin{thebibliography}{99}
\bibitem{ref1} See, for example, C. Giunti and C. Kim, {\it Fundamentals of Neutrino Physics and Astrophysics} (Oxford University Press, Oxford, 2007).

\bibitem{ref2} B. Kayser, {\it Proceedings of the SLAC Summer Institute of 2004}, eConf {\bf C040802}, L004 (2004); arXiv: hep-ph/0506165.

\bibitem{ref3} B. Kayser, {\it Proceedings of the 9th Int. Conf. on Flavor Physics and CP Violation}, Ed. A. Soffer, eConf C110523; arXiv: 1110.3047. What follows is based on this reference, and quotes a bit of its text.

\bibitem{ref4} This treatment of a problem involving particle mixing follows a method described in B. Kayser and L. Stodolsky, \plb{359}{343}{1995}.

\bibitem{ref5} This was first noticed by E. Akhmedov and A. Smirnov, {\it Found. Phys.} {\bf 41}, 1279 (2011); arXiv: 1008.2077.

\bibitem{ref6} B. Kayser, \prd{24}{110}{1981}.

\bibitem{ref7} S. Davidson, E. Nardi, and Y. Nir, {\it Phys. Rept.} {\bf 466}, 105 (2008); arXiv: 0802.2962.

\bibitem{ref8} B. Kayser, {\it Proceedings of the 22nd Rencontres de Blois}, Eds. L. Celniker, J. Dumarchez, B. Klima, and J. Tr\^{a}n Thanh Van (Gioi Publishers, Vietnam, 2011), p. 91; arXiv: 1012.4469.

\bibitem{ref9} J. Bahcall, {\it Neutrino Astrophysics} (Cambridge University Press, Cambridge, 1989).

\bibitem{ref10} M. Fukugita and T. Yanagida, {\it Physics of Neutrinos and Applications to Astrophysics} (Springer, Berlin, 2003).

\bibitem{ref11} S. Elliott and P. Vogel, {\it Ann. Rev. Nucl. Part. Sci.} {\bf 52}, 115 (2002); arXiv: hep-ph/0202264.

\bibitem{ref12} W. Rodejohann, {\it Int. J. Mod. Phys.} {\bf E20}, 1833 (2011); arXiv: 1106.1334.

\bibitem{ref13} B.Kayser, {\it Proceedings of the Carolina International Symposium on Neutrino Physics}, Eds. F. Avignone, R. Creswick, K. Kubodera, and M. Purohit (IOP Publishing, Bristol, 2009), p. 012013; arXiv: 0903.0899

\bibitem{ref14} Particle Data Group (K. Nakamura, {\it et al.}), {\it J. Phys. G }{\bf 37}, 1 (2010), accessible at pdg.lbl.gov, and its periodic updates.


\end{thebibliography}
\end{document}